\documentclass[%
reprint, 
amsmath,amssymb, 
aps,prl]
{revtex4-2}

\usepackage[utf8]{inputenc}
\usepackage{graphicx}
\usepackage{dcolumn}
\usepackage{bm}
\usepackage{color}

\def\d{{\rm d}}   	
\def\i{{\rm i}}		

\begin{document}

\title{K\"ahler geometry of black holes and gravitational instantons}%

\author{Steffen Aksteiner}
 \email{\texttt{steffen.aksteiner@aei.mpg.de}}

\author{Bernardo Araneda}%
 \email{\texttt{bernardo.araneda@aei.mpg.de}}

\affiliation{%
Max Planck Institute for Gravitational Physics (Albert Einstein Institute), Am M\"uhlenberg 1, D-14476 Potsdam, Germany 
 }%

\date{\today}

\begin{abstract}
We obtain a closed formula for the K\"ahler potential of a broad class of 
four-dimensional Lorentzian or Euclidean conformal ``K\"ahler'' 
geometries, including the Pleba\'nski-Demia\'nski class and 
various gravitational instantons such as Fubini-Study and Chen-Teo.
We show that the 
K\"ahler potentials of Schwarzschild and Kerr 
are related by a Newman-Janis shift. 
Our method also shows that a class of supergravity black holes, 
including the Kerr-Sen spacetime, is Hermitian 
(but not conformal K\"ahler).
We finally show that the integrability conditions of complex structures
lead naturally to the (non-linear) Weyl double copy, and we give new 
vacuum and non-vacuum examples of this relation.
\end{abstract}


\maketitle

\section{Introduction}

Complex methods as a tool to investigate spacetime structure 
in General Relativity (GR) have a long and fruitful history 
of remarkable developments.
Profound constructions, pioneered by
Penrose, Newman, Pleba\'nski, Robinson, Trautman, \cite{Penrose76, Newman76, Plebanski75, PlebanskiRobinson, 
Trautman62, NewmanJanis} among others, include twistor theory and 
heavenly structures,
but there are also simple yet intriguing results such as the 
``Newman-Janis shift'' relating special solutions
via complex coordinate transformations.

An important insight regarding complex structures in GR  
is provided by Flaherty \cite{Flaherty0}, who showed that 
type D vacuum and Einstein-Maxwell spacetimes possess 
an analogue of the Hermitian structures of Riemannian geometry. 
In Lorentz signature, 
a Hermitian structure must necessarily be 
complex-valued, so its integrability properties are more subtle 
than in the Euclidean case. 
Flaherty gave a comprehensive analysis of such properties \cite{Flaherty, FlahertyHeld},
and he found that the above classes of type D spacetimes are not only Hermitian but also satisfy the Lorentzian analogue of the conformal K\"ahler condition. 

In Riemannian geometry, K\"ahler metrics are encoded in 
``generating functions'' or scalar K\"ahler potentials. 
An analogous feature in GR occurs in perturbation theory,
where perturbative fields are generated by scalar Debye potentials. 
These potentials are instrumental for modern 
studies of e.g. black hole stability and gravitational wave physics.
The increasing interest in nonperturbative structures for gravitational 
wave science \cite{Buonanno}, together with the importance of scalar 
potentials for perturbation theory, motivate the question of whether there 
are ``Debye potentials'' for exact, 
astrophysically relevant solutions of GR. 
Moreover, the recently discovered applications of the Newman-Janis 
shift \cite{ArkaniHamed, Guevara} suggest that complex structures 
in GR may play an important role in the understanding of 
such nonperturbative structures.

Motivated by the above considerations, in this paper we develop a 
method to find the K\"ahler potentials of a broad class of 
geometries, including black holes and gravitational instantons, 
and we show intimate connections of this approach 
with other theoretical structures of modern interest 
for gravitational wave physics 
such as the Newman-Janis shift and the double copy relation 
between gauge and gravity theories.

As an example, consider the Kerr metric $g$
with parameters $M, a$ for mass and angular momentum per mass, respectively. In Boyer-Lindquist coordinates $(t,r,\theta,\phi)$, 
the metric is block diagonal in $(\d t,\d \phi)$ and $(\d r, \d \theta)$, with components in the first block given by
\begin{align*}
g_{tt} ={}& \left( \Delta - a^2 \sin^2\theta \right)\Sigma^{-1}, \\
g_{t\phi} ={}&  -a(\Delta - (r^2 + a^2))\Sigma^{-1} \sin^2\theta, \\
g_{\phi\phi} ={}& (a^2 \Delta \sin^2\theta  - (r^2 + a^2)^2)\Sigma^{-1} \sin^2\theta,
\end{align*}
where $\Sigma = r^2 + a^2 \cos^2\theta, \Delta = r^2 - 2 M r  + a^2$. 
Following Flaherty \cite{Flaherty}, one can find four complex 
scalar fields $(z^{0},z^{1},\tilde{z}^{0},\tilde{z}^{1})$, 
defined by 
\begin{align*} 
\d z^{0}={}& \d t  -  (a^2 + r^2) \Delta^{-1} \d r - \i a \sin\theta \d\theta,\\
\d z^{1} ={}& \d\phi - a \Delta^{-1} \d r -  \i \csc\theta \d\theta, \\
\d\tilde{z}^{0} ={}& \d t + (a^2 + r^2) \Delta^{-1} \d r + \i a \sin\theta \d\theta,\\
\d\tilde{z}^{1} ={}& \d\phi + a \Delta^{-1} \d r + \i \csc\theta \d\theta,
\end{align*}
such that the Kerr metric is 
\begin{align*}
g ={}&g_{tt} \, \d z^{0} \d\tilde{z}^{0}
 + g_{t\phi} \, (\d z^{0} \d\tilde{z}^{1} + \d z^{1} \d\tilde{z}^{0})
 + g_{\phi\phi} \, \d z^{1} \d\tilde{z}^{1}.
\end{align*}
In addition, letting $\Omega^{-2} = (r - \i a \cos\theta)^2$, 
there must exist a scalar $K$ such that 
$ g_{tt} ={} \Omega^{-2} \partial_{z^{0}}\partial_{\tilde{z}^{0}} K$, 
$g_{t\phi} ={} \Omega^{-2} \partial_{z^{0}}\partial_{\tilde{z}^{1}} K$, 
$g_{\phi\phi} ={} \Omega^{-2} \partial_{z^{1}}\partial_{\tilde{z}^{1}} K$.
However, expressions for $K$ do not seem to have been obtained in 
the literature.

The method developed in this paper computes the generating 
function/K\"ahler potential of Kerr to be 
\begin{align*}
 K = 4\int r \Delta^{-1} {\rm d}r + 4 \log\sin\theta. 
\end{align*}
As a function $K=K(z,\tilde{z})$, the potential 
fully generates the spacetime geometry. 
Moreover, using K\"ahler transformations,
we shall show that the K\"ahler potentials for Kerr and 
Schwarzschild are simply related by a Newman-Janis shift. 
More generally, the geometries studied in this paper 
include the general Pleba\'nski-Demia\'nski class \cite{PD} 
as well as the Chen-Teo family \cite{ChenTeo} of gravitational instantons. 
Our method also allows us to prove that a general class of supergravity black holes \cite{Chow}, including the Kerr-Sen spacetime \cite{Sen}, 
has a Hermitian (not conformal K\"ahler) structure.

In addition, we shall show that the integrability of complex structures leads to the existence of special scalar and massless free fields associated to the geometry, that can be combined to give a unified geometric description of the `Weyl double copy' 
\cite{BHDC,typeDDC, typeNDC}. 
Our results contain not only the type D and N double copies, but also provide new examples of this relation for both vacuum and non-vacuum geometries, including e.g. the general Einstein-Maxwell Pleba\'nski-Demia\'nski class and the Fubini-Study and Chen-Teo instantons.

Importantly, we shall not impose any field equations: the conformal 
K\"ahler property of the geometries we study does not depend on 
a particular field theory. 
This means that the conformal factor does not, in principle, play a role 
in our construction, but we shall nevertheless include it since 
it arises naturally in GR, where the Einstein and K\"ahler metrics are 
conformally related.

\section{Complexified K\"ahler geometry}

Given a 4-dimensional complex geometry $(M,g)$,
we define an almost-Hermitian structure
\footnote{The standard definition requires $J$ to be real-valued, 
but we allow $J$ to be complex-valued since we also include 
metrics with Lorentzian signature. 
We shall continue to use terminology, such as Hermitian, 
Kaehler, etc., as in the Riemannian case.}
as a $(1,1)$ tensor field $J$ such that $J^{2}=-\mathbb{I}$
and $g(J\cdot,J\cdot)=g(\cdot,\cdot)$. 
The tangent bundle decomposes as $TM=T^{+}\oplus T^{-}$, 
where $T^{\pm}$ corresponds to vectors with eigenvalue 
$\pm{\rm i}$ under $J$. 
We say that the almost-Hermitian structure is integrable,
and is thus a Hermitian structure,
if $[T^{\pm},T^{\pm}]\subset T^{\pm}$ (for {\em both} signs), 
where $[\cdot,\cdot]$ is the Lie bracket of vector fields. 
One can show that a Hermitian structure implies that there are  
four complex scalars $(z^{i},\tilde{z}^{i})$ such that
\begin{align}
g = g_{i\tilde{j}}{\rm d}z^{i}{\rm d}\tilde{z}^{j},
  \label{HermitianMetric}
\end{align}
where $g_{i\tilde{j}}=g(\partial_{z^{i}},\partial_{\tilde{z}^{j}})$, 
with $i=0,1$, $\tilde{j}=\tilde{0},\tilde{1}$.

The fundamental 2-form is defined by 
$\kappa(\cdot,\cdot):=g(J\cdot,\cdot)$. 
We say that a Hermitian geometry is K\"ahler if ${\rm d}\kappa=0$, 
and conformal K\"ahler if there is a scalar field $\Omega^{2}$ 
such that ${\rm d}\hat\kappa=0$, where 
$\hat\kappa=\Omega^{2}\kappa$.
By the complex version of the Poincar\'e Lemma, 
if ${\rm d}\hat\kappa=0$ then there exists, locally, a complex 
scalar $K$ such that
\begin{align}
 \hat{g}_{i\tilde{j}} = \partial_{z^i}\partial_{\tilde{z}^j}K
\label{ddK}
\end{align}
where $\hat{g}_{i\tilde{j}}=\Omega^{2}g_{i\tilde{j}}$,
cf. \cite[Theorem IX.8]{Flaherty}.
We say that $K$ is a K\"ahler potential. It is not unique: 
one has the freedom to perform {\em K\"ahler transformations}
\begin{align}
 K \to K + F(z^{i})+\tilde{F}(\tilde{z}^{i}). 
 \label{KahlerTransformations}
\end{align}

The K\"ahler potential can be found by integrating $\eqref{ddK}$.
Define $p_{j}:=\partial K/ \partial\tilde{z}^{j}$, then 
$p_{j}=\int\hat{g}_{i\tilde{j}}{\rm d}z^{i}$. 
Integrating once again, the potential is  
$K=\int p_{i}{\rm d}\tilde{z}^{i}$.

In this work we shall study geometries whose metric has the 
block-diagonal form
\begin{align}
 g =  a_{ij}{\rm d}\sigma^{i}{\rm d}\sigma^{j} 
 + b_{IJ}{\rm d}x^{I}{\rm d}x^{J}
 \label{SpecialClass}
\end{align}
for some coordinates $\sigma^{i}=(\tau,\phi)$ and $x^{I}=(x,y)$, 
and known functions $a_{ij},b_{IJ}$.
Introduce an orthonormal coframe 
$e^{1}=c^{1}_{i}{\rm d}\sigma^{i}$, 
$e^{2}=c^{2}_{I}{\rm d}x^{I}$, 
$e^{3}=c^{3}_{I}{\rm d}x^{I}$, 
$e^{4}=c^{4}_{i}{\rm d}\sigma^{i}$, 
for some functions $c^{1}_{i},c^{2}_{I},c^{3}_{I},c^{4}_{i}$; 
such that $g=e^{1}\otimes e^{1}+...+e^{4}\otimes e^{4}$.
Define now a null coframe by 
\begin{equation}\label{NullFrame}
\begin{aligned}
 \ell ={}& \tfrac{1}{\sqrt{2}}(e^{1} + {\rm i}e^{2}), \qquad
 n = \tfrac{1}{\sqrt{2}}(e^{1} - {\rm i}e^{2}), \\
 m ={}& \tfrac{1}{\sqrt{2}}(e^{3} + {\rm i}e^{4}), \qquad
 \tilde{m} = \tfrac{1}{\sqrt{2}}(-e^{3} + {\rm i}e^{4}).
\end{aligned}
\end{equation}
The metric is $g=2(\ell\odot n - m\odot\tilde{m})$.
We shall consider almost-Hermitian structures whose fundamental 
2-forms are $\kappa_{\pm}=\i(\ell\wedge n \pm m\wedge\tilde{m})$.
For concreteness, let us focus on $\kappa_{-}\equiv\kappa$.

Let $V^{i}=(-\partial_{\phi},\partial_{\tau})$.
We define the 1-forms
\begin{align}
 \omega^{i}:=\mu V^{i}\lrcorner \ (\ell\wedge m), \qquad 
 \tilde\omega^{i}:=\tilde{\mu} V^{i}\lrcorner \ (n\wedge\tilde{m})
\end{align}
where
$\mu^{-1}=(\ell\wedge m)(\partial_{\tau},\partial_{\phi})$ and 
$\tilde\mu^{-1}=(n\wedge \tilde{m})(\partial_{\tau},\partial_{\phi})$.
A calculation shows that
\begin{align}\label{omegas}
 \omega^{i} = {\rm d}\sigma^{i} + E^{i}_{I}{\rm d}x^{I}, \qquad
 \tilde\omega^{i} = {\rm d}\sigma^{i} - E^{i}_{I}{\rm d}x^{I},
\end{align}
for some functions $E^{i}_{I}=\omega^{i}(\partial_{I})$, 
where $\partial_{I}=\partial/\partial x^{I}$.
In addition, the metric and fundamental 2-form are
\begin{align}\label{gk}
 g = g_{i\tilde{j}} \ \omega^{i}\odot\tilde{\omega}^{j}, 
 \qquad 
 \kappa = \tfrac{\i}{2}g_{i\tilde{j}} \  \omega^{i}\wedge\tilde{\omega}^{j}
\end{align}
where $g_{i\tilde{j}}=g(\partial_{\sigma^i},\partial_{\sigma^j})$.
Note that this implies that
$g_{0\tilde{0}}=g_{\tau\tau}$, 
$g_{0\tilde{1}}=g_{1\tilde{0}}=g_{\tau\phi}$, 
$g_{1\tilde{1}}=g_{\phi\phi}$. 

The almost-Hermitian structure is integrable iff
${\rm d}\omega^{i}=0={\rm d}\tilde\omega^{i}$: if this is satisfied, then
there will be (locally) $z^{i},\tilde{z}^{i}$ such that 
$\omega^{i}={\rm d}z^{i}$ and $\tilde\omega^{i}={\rm d}\tilde{z}^{i}$, 
and from the first equation in \eqref{gk} we see that 
the metric \eqref{SpecialClass} will have the Hermitian 
expression \eqref{HermitianMetric}.
Using \eqref{omegas}, this integrability condition has a simple 
form: ${\rm d}\omega^{i}=0={\rm d}\tilde\omega^{i}$ iff
\begin{align}\label{ConditionIntegrability}
 E^{i}_{I} = E^{i}_{I}(x^{J}) \quad\text{and}\quad  \partial_{[I}E^{i}_{J]}=0.
\end{align}
The second equation implies 
that locally, there are functions $\psi^{0}(x^{I}),\psi^{1}(x^{I})$ 
such that $E^{i}_{I}=\partial_{I}\psi^{i}$.
The $(z^{i},\tilde{z}^{i})$ coordinates will then be given by
\begin{align}\label{zCoordinates}
 z^{i} = \sigma^{i} + \psi^{i}, \qquad \tilde{z}^{i} = \sigma^{i} - \psi^{i}.
\end{align}
The associated vector fields are 
$\partial_{z^i}=\frac{1}{2}(\partial_{\sigma^i}+\partial_{\psi^i})$, 
$\partial_{\tilde{z}^i}=\frac{1}{2}(\partial_{\sigma^i}-\partial_{\psi^i})$.
In terms of $x^I$, we have 
$\partial_{\psi^i}=E^{I}_{i}\partial_{I}$, where $E^{I}_{i}$ 
is the inverse of $E^{i}_{I}$ (thought of as a $2\times2$ matrix).

We shall now assume that $\partial_{\sigma^1}=\partial_{\tau}$ and 
$\partial_{\sigma^2}=\partial_{\phi}$ are Killing vectors.
This includes all of the examples studied in this paper. 
Using 
$\kappa=-{\rm i}g_{i\tilde{j}}{\rm d}\sigma^{i}\wedge{\rm d}\psi^{j}$ 
and $\hat{g}_{i\tilde{j}}:=\Omega^{2}g_{i\tilde{j}}$, 
a short calculation shows that the conformal K\"ahler condition 
${\rm d}(\Omega^{2}\kappa)=0$ is
\begin{equation}\label{ConformalKahlerCondition}
\begin{aligned}
 \partial_{\psi^{0}}\hat{g}_{\tau\phi} 
 - \partial_{\psi^{1}}\hat{g}_{\tau\tau} = 0 =
  \partial_{\psi^{0}}\hat{g}_{\phi\phi} 
 - \partial_{\psi^{1}}\hat{g}_{\tau\phi}.
\end{aligned}
\end{equation}

Assuming the above conditions, the formula 
$K=\int p_{i}{\rm d}\tilde{z}^{i}$ for the K\"ahler potential can be 
rewritten as follows. 
From Cartan's formula 
$\pounds_{v}\hat\kappa = {\rm d}(v\lrcorner\hat\kappa) 
 + v\lrcorner{\rm d}\hat\kappa$, we deduce that the 
Killing fields have Hamiltonians, i.e. functions $H_0, H_{1}$ such that
${\rm d}H_{i}=\partial_{\sigma^{i}}\lrcorner \ \hat\kappa
 = -{\rm i}\hat{g}_{i\tilde{j}}{\rm d}\psi^{j}$,
where the second equality follows from the expression of  
$\hat\kappa$ in terms of $\sigma^i,\psi^{i}$.
Choosing $K$ to be independent of $\sigma^i$, we get
\begin{align}
 K = -4{\rm i} \int H_{i}{\rm d}\psi^{i}. \label{KahlerPotentialGeneral}
\end{align}
The integration in $\psi^{i}$ can be replaced by an integration in 
$x^{I}$ by using ${\rm d}\psi^{i}=E^{i}_{I}{\rm d}x^{I}$.

To recover real metrics with different signatures, we impose 
reality conditions on the null coframe $(\ell,n,m,\tilde{m})$.
Euclidean signature $(++++)$ corresponds to requiring 
$n=\bar{\ell}$ and $\tilde{m}=-\bar{m}$. 
The functions $E^{i}_{I}$ in \eqref{omegas} are then purely imaginary, 
so $\tilde\omega^{i}=\bar\omega^{i}$ and $\tilde{z}^{i}=\bar{z}^{i}$.
Lorentzian signature $(+---)$ corresponds to $\ell,n$ real and 
$\tilde{m}=\bar{m}$.
The functions $E^{i}_{I}$ in \eqref{omegas} are generally complex, 
so $z^{i}$ and $\tilde{z}^{i}$ in \eqref{zCoordinates} are not complex 
conjugates.

\section{Black holes and instantons} 

\paragraph{Diagonal metrics.}
Consider the special case of \eqref{SpecialClass} where 
$g=g_{\tau\tau}{\rm d}\tau^{2}+g_{\phi\phi}{\rm d}\phi^{2}
+g_{xx}{\rm d}x^{2}+g_{yy}{\rm d}y^{2}$.
We choose the frame such that the functions $E^{i}_{I}$ in 
\eqref{omegas} 
are $E^{\tau}_{x} = {\rm i} \sqrt{g_{xx}/g_{\tau\tau}}$, $E^{\tau}_{y}=0$, 
$E^{\phi}_{x}=0$, $E^{\phi}_{y}=-{\rm i}\sqrt{g_{yy}/g_{\phi\phi}}$.
The Hermitian condition is equivalent to 
$\partial_{y}(g_{xx}/g_{\tau\tau})=0$, 
$\partial_{x}(g_{yy}/g_{\phi\phi})=0$, 
and the conformal K\"ahler condition is 
$\partial_{x}(\Omega^{2}g_{\tau\tau})=0$, 
$\partial_{y}(\Omega^{2}g_{\phi\phi})=0$. 

A simple example is an arbitrary static, spherically symmetric spacetime 
$g=f(r)\d\tau^2 - h(r)\d r^2 - r^2(\d\theta^2+\sin^2\theta\d\phi^2)$. 
Using $\Omega^{2}=1/r^2$, and {\em regardless} of the form of 
$f(r),h(r)$, the geometry is conformal K\"ahler. 
This includes not only the well-known 
spherical black hole spacetimes but also 
solutions from the Einstein-scalar field system such as 
the Janis-Newman-Winicour wormhole \cite{JNW}.
In the special case $h=f^{-1}$, the K\"ahler potential is 
given by $K = 4\left[\int[rf(r)]^{-1}\d r + \log\sin\theta \right]$.

\paragraph{The Pleba\'nski-Demia\'nski class.}
Consider the metric \eqref{SpecialClass} with 
\begin{equation}
\begin{aligned}\label{PDfamily}
 g_{\tau\tau} ={}& [\Delta_r - a^2 \Delta_x] / (\Pi \Sigma), \\
 g_{\tau\phi}  ={}& 
  a[(r^2 + a^2) \Delta_x - (1 - x^2)\Delta_r]/ (\Pi \Sigma),\\
 g_{\phi\phi} ={}& [a^2 (1 - x^2)^2 \Delta_r - (r^2 + a^2)^2\Delta_x]
 /(\Pi \Sigma), \\
 g_{xx} ={}& -\Sigma/(\Pi\Delta_{x}), \quad 
 g_{xy}=0, \quad g_{yy} = -\Sigma/(\Pi\Delta_{r})
\end{aligned}
\end{equation}
where $y\equiv r$, $\Sigma = r^2 + a^2 x^2$, 
$\alpha$ and $a$ are constants, 
and $\Pi = \Pi(r,x)$, $\Delta_{x}=\Delta_{x}(x)$, $\Delta_{r}=\Delta_{r}(r)$
are arbitrary functions of their arguments.
We find that, {\em regardless} of the 
specific form of $\Pi, \Delta_{r},\Delta_{x}$,
the geometry is conformal K\"ahler, with
complex coordinates 
\begin{equation}\label{zCoordPD}
\begin{aligned}
 z^{0}={}& \tau - (r^{*}  - {\rm i}a x^{*}), & 
 z^{1} ={}& \phi - (ar^{\sharp} - {\rm i}x^{\sharp}), \\
 \tilde{z}^{0} ={}& \tau + (r^{*} -{\rm i}a x^{*}), &
 \tilde{z}^{1} ={}& \phi + (a r^{\sharp} - {\rm i}x^{\sharp}),
\end{aligned}
\end{equation}
where $r^{*},x^{*},r^{\sharp},x^{\sharp}$ are defined by
\begin{equation}\label{StarAndSharp}
\begin{aligned}
 {\rm d}r^{*}={}& (r^2 + a^2)\Delta_r^{-1}{\rm d}r, &
 {\rm d}x^{*} ={}& (1 - x^2)\Delta_x^{-1}{\rm d}x, \\
 {\rm d}r^{\sharp} ={}& \Delta_r^{-1} {\rm d}r, &
 {\rm d}x^{\sharp} ={}& \Delta_x^{-1} {\rm d}x,
\end{aligned}
\end{equation}
and the conformal factor is
\begin{align}
\Omega^{2}=\Pi/(r-{\rm i}ax)^{2}. \label{ConformalFactor}
\end{align}
The K\"ahler form $\hat\kappa=\Omega^{2}\kappa$ is given by
\begin{align}
\nonumber \hat\kappa = \frac{{\rm i}}{(r-{\rm i}ax)^{2}} & \left\{ 
 -\d\phi\wedge[a(1-x^2)\d r - \i(r^2+a^2)\d x] \right. \\
 & \left. + \d\tau\wedge(\d r - \i a\d x)
\right\}.
 \label{MaxwellField}
\end{align}
Notice that this is independent of $\Delta_{r}, \Delta_{x}$.
The Hamiltonians are
$H_{0}=-\i/(r - \i ax)$ and $H_{1}=\i(a+\i rx)/(r - \i ax)$,
hence, using \eqref{KahlerPotentialGeneral}, 
we find that the K\"ahler potential is
\begin{align}
 K = 4\left[ \int\frac{r}{\Delta_{r}}{\rm d}r 
 - \int\frac{x}{\Delta_{x}}{\rm d}x  \right]. 
 \label{KahlerPotentialPD}
\end{align}
We stress that the existence of this potential is independent of 
the explicit form of the functions $\Delta_{r},\Delta_{x}$.

The Pleba\'nski-Demia\'nski family \cite{PD, Podolsky} 
is \eqref{PDfamily} with $\Pi = (1- \alpha r x)^2$ and 
\begin{equation*}
\begin{aligned}
\Delta_x={}&1 + \tfrac{2 N }{a}x -  x^2 + 2 \alpha M x^3 
 - \bigl(\tfrac{\lambda}{3} a^2  + \alpha^2 (Q^2 + a^2)\bigr) x^4,\\
\Delta_{r}{}={}& Q^2 + a^2 - 2 M r + r^2 
 - \tfrac{2 \alpha N }{a}r^3 -  (\alpha^2 + \tfrac{1}{3} \lambda) r^4,
\end{aligned}
\end{equation*}
where $Q^{2}=q_{e}^2 + q_{m}^2$, and $\lambda,q_{e},q_{m}$ 
correspond, respectively, to cosmological constant and
electric and magnetic charges. The rest of the parameters can 
be related to mass, angular momentum, acceleration, and 
NUT charge, cf. \cite{Podolsky} for details.
This is the general type D solution (assuming non-null orbits of the isometry group)
of the Einstein equations 
with an aligned electromagnetic field.

We note that, for the case $Q=0$, the 
transformation $(r,M) \leftrightarrow \pm(\i ax, \i N)$ leaves 
the K\"ahler potential and the metric invariant, 
and the coordinates \eqref{zCoordPD} change according to 
$z^{i}\leftrightarrow \tilde{z}^{i}$ for $+$ and are invariant for $-$. 
A detailed analysis of this and other dualities will 
be given in a separate work \cite{SBToappear}.

\paragraph{Newman-Janis shifts.}
For the Schwarzschild and Kerr spacetimes (putting $x=\cos\theta$), 
we find the K\"ahler potentials to be
\begin{subequations}\label{KahlerBHs}
\begin{align}
 K_{\rm schw} ={}& 4\log((r-2M)\sin\theta), 
 \label{KahlerSchw0} \\
\nonumber K_{\rm kerr} ={}& 4\left[ \frac{1}{2}\log(r^{2}-2Mr+a^{2}) + 
 \log\sin\theta \right. \\
 & \left. -\frac{M}{\sqrt{M^{2}-a^{2}}}
 \tanh^{-1}\left( \frac{r-M}{\sqrt{M^{2}-a^{2}}} \right) \right], 
 \label{KahlerKerr0}
\end{align}
\end{subequations}
where we assume the non-extreme case $M^{2}>a^{2}$. 
Using \eqref{zCoordPD} and K\"ahler transformations 
\eqref{KahlerTransformations}, a calculation shows that 
\eqref{KahlerSchw0} and \eqref{KahlerKerr0} are equivalent to 
\begin{subequations}
\begin{align}
 K_{\rm schw} ={}& 4\left[ -\frac{r}{2M} + \log\sin\theta \right], 
 \label{KahlerSchw} \\
 K_{\rm kerr} = 
 {}& 4\left[-\frac{(r-{\rm i}a\cos\theta)}{2M} + \log\sin\theta
 \right] \label{KahlerKerr}
\end{align}
\end{subequations}
where we assume $M\neq 0$. 
Thus, the K\"ahler potentials are related by a Newman-Janis shift 
$r \to r - {\rm i}a\cos\theta$ \cite{NewmanJanis}, 
although it is not at all obvious from \eqref{KahlerBHs}.

For $M=0$, which corresponds (locally) to flat spacetime,
we can see the Newman-Janis shift as follows. 
Consider complexified Minkowski space, 
in complexified spherical coordinates $(r_{c},\theta_{c},\phi_{c})$. 
In terms of complexified inertial coordinates $(t_{c},x_{c},y_{c},z_{c})$,
we have the usual relations 
$x^{2}_{c}+y^{2}_{c}=r^{2}_{c}\sin^{2}\theta_{c}$, 
$z_{c}=r_{c}\cos\theta_{c}$.
The K\"ahler potential can be shown to be $K=4\log(r_{c}\sin\theta_{c})$. 
Consider first the real slice $\mathbb{M}$ given by 
$\{t_{c}=t, \ x_{c}=x, \ y_{c}=y, \ z_{c}=z \}$, where $t,x,y,z$ are real. 
Then $(r_{c},\theta_{c},\phi_{c})$ become ordinary 
real spherical coordinates, and the K\"ahler potential is 
\begin{align}
 K |_{\mathbb{M}} = {}& 4\log(r\sin\theta). \label{KSM0}
\end{align}
Now consider a different real slice $\mathbb{M}'$ given by 
a Newman-Janis shift \cite{Newman2002}:
$ \{t_{c}=t, \ x_{c}=x, \ y_{c}=y, \ z_{c}=z - \i a\}$, where 
$a$ is a real constant.
Choosing the complex radius to be 
$r_{c}=r-\i a\cos\theta$, a calculation gives 
$x^2+y^2=(r^{2}+a^{2})\sin^2\theta$, so
\begin{align}
 K |_{\mathbb{M}'} = {}& 4\left[ \tfrac{1}{2}\log(r^{2}+a^{2}) + 
 \log\sin\theta \right]. \label{KKM0}
\end{align}
Eqs. \eqref{KSM0} and \eqref{KKM0} correspond, respectively, 
to the $M\to 0$ limits in \eqref{KahlerSchw0} and \eqref{KahlerKerr0}.

\paragraph{Supergravity black holes.}
Consider the metric \eqref{SpecialClass} with
\begin{equation}\label{NonExample}
\begin{aligned}
 g_{\tau\tau} ={}& (R-U)/W, \quad g_{\tau\phi} = (RW_{u}+UW_{r})/W, \\
 g_{\phi\phi} ={}& (RW^{2}_{u}-UW^{2}_{r})/W, \\
 g_{xx}={}&-W/R, \quad g_{xy}=0, \quad g_{yy} = -W/U,
\end{aligned}
\end{equation}
where $x\equiv r$, $y\equiv u$, $(R,W_{r})$ and $(U,W_{u})$ 
are arbitrary functions of $r$ and $u$ respectively,
and $W=a(W_{r}+W_{u})$, with  real constant $a$.
The metric \eqref{NonExample} includes a general class of 
black hole solutions of supergravity \cite{Chow}, 
in particular the Kerr-Sen black hole \cite{Sen}.

Using the almost-Hermitian structure associated to the frame 
given in \cite[Eq. (4.79)]{Chow},
our method shows that the geometry 
\eqref{NonExample} is Hermitian, with complex coordinates  
\eqref{zCoordinates}, where $\psi^{0}=r^{*} + \i u^{*}$, 
$\psi^{1}=r^{\sharp} - \i u^{\sharp}$, and 
$\d r^{*}=a(W_r/R)\d r$, $\d u^{*}= a(W_u/U)\d u$, 
$\d r^{\sharp} = (a/R)\d r$, $\d u^{\sharp} = (a/U)\d u$.
However, the conformal K\"ahler condition 
\eqref{ConformalKahlerCondition} does not hold for this 
Hermitian structure.

\paragraph{Gravitational instantons.}
We now specialize to Euclidean signature.
Consider first the metric \eqref{SpecialClass} with
\begin{align*}
 & g_{\tau\tau}=\tfrac{a^2x^2}{4(1+x^2)^2},  
 \quad g_{\phi\phi}=g_{\tau\tau}(1+x^2\sin^2 y), \quad g_{xy}=0, \\
 & g_{\tau\phi}=g_{\tau\tau}\cos y, \quad 
 g_{xx}=\tfrac{4}{a^{2}x^2}g_{\tau\tau},
  \quad g_{yy}= (1+x^2)g_{\tau\tau},
\end{align*}
where $a$ is an arbitrary constant. 
Using ``$(\mp)$'' to denote quantities associated to $\kappa_{\mp}$,
one can choose frames such that the functions in \eqref{omegas} are 
$E^{\tau}_{(\mp)x}= 2\i/(ax)$, $E^{\tau}_{(\mp)y}=\pm \i\cot y$, 
$E^{\phi}_{(\mp)x}=0$,  $E^{\phi}_{(\mp)y}=\mp\i\csc y$.
Then a calculation shows that the geometry is conformal K\"ahler 
w.r.t. both sides, with $\Omega^{2}_{\mp}=[(1+x^2)/x^2]^{1\mp 1/a}$.
For $a=\pm1$, one side becomes K\"ahler and the metric is Einstein:  
this is the Fubini-Study metric in $\mathbb{CP}^{2}$.

A new family of gravitational instantons was 
discovered by Chen and Teo \cite{ChenTeo}. 
This is a toric, Ricci-flat geometry of the form \eqref{SpecialClass}
that depends on seven parameters $k,\nu,a_{0}\dots a_{4}$. 
The non-trivial metric components are
$g_{\tau\tau}, g_{\tau\phi}, g_{\phi\phi}, g_{xx}, g_{yy}$, 
and depend on functions $F,G,H,X,Y$ given explicitly 
in \cite[Eq. (2.1)]{ChenTeo}. 
The family contains other known instantons such as Eguchi-Hanson 
and Euclidean Pleba\'nski-Demia\'nski.
It was recently shown \cite{Aksteiner} that the Chen-Teo geometry is 
one-sided type D, and thus (from Ricci-flatness) conformal K\"ahler, 
with $\Omega^2 = (x-y)^2/(\nu x+y)^2$. 
Our method computes the complex coordinates to be  
$\d z^0 = \d\tau+\d\psi^0$, 
$\d z^1 = \d\phi + \d\psi^1$ (cf. \eqref{zCoordinates}), 
where $\d\psi^0 = E^{\tau}_{x}\d x + E^{\tau}_{y}\d y$, 
$\d\psi^1 = E^{\phi}_{x}\d x + E^{\phi}_{y}\d y$, and
\begin{equation}
\begin{aligned}
 E^{\tau}_{x} ={}& \i\frac{\sqrt{k}}{F} \left( \frac{Gx}{X}+\frac{Hy}{(x-y)} \right), 
 & E^{\phi}_{x}={}& \frac{-\i\sqrt{k}x}{X}, \\
 E^{\tau}_{y} ={}&  -\i\frac{\sqrt{k}}{F} \left( \frac{Gy}{Y}+\frac{Hx}{(x-y)} \right), 
 & E^{\phi}_{y}={}& \frac{\i\sqrt{k}y}{Y}.
\end{aligned}
\end{equation}
The Hermitian condition \eqref{ConditionIntegrability} reduces 
to $\partial_{y}E^{\tau}_{x}-\partial_{x}E^{\tau}_{y}=0$, 
which provides an interpretation for eq. (3.49) in \cite{Aksteiner}.
The Hamiltonians are
\begin{equation*}
 H_{0} = \frac{\sqrt{k}(x-y)}{(1+\nu)(\nu x+y)}, \quad
 H_{1} = \frac{\sqrt{k}f(x,y)}{(\nu x+y)(x-y)},
\end{equation*}
where $f(x,y)=(\nu-1)(a_{0}+a_4x^2y^2) - a_2 x(\nu(x-2y)+y)
+ (a_1+a_3 xy)(\nu x-y) $.
The K\"ahler potential can now be computed using 
\eqref{KahlerPotentialGeneral}:
\begin{align*}
K = \frac{4k}{1+\nu} \left[ 4 (1-\nu) \log(x-y) -  \int \frac{h_{1}}{X} \d x + \int \frac{h_{2}}{Y} \d y \right]
\end{align*}
where 
$h_{1}(x) = a_{1}{} + a_{2}{} (1 - 2 \nu)x + a_{3}{} (2 - \nu ) x^2 + 2 a_{4}{} (1 - \nu ) x^3  $ and
$h_{2}(y) = a_{1}{} \nu  - a_{2}{} y - a_{3}{} (1 - 2 \nu ) y^2 - 2 a_{4}{} (1 - \nu ) y^3$.

\section{Double copy structures}

In string theory, the KLT relations \cite{KLT} imply that gravitational 
amplitudes are closely related to the square of Yang-Mills amplitudes. 
The extension of these relations to field theory is known 
as the `double copy'. 
At the classical level, a recent formulation is 
the `curved Weyl double copy' \cite{typeDDC}, which asserts
that for some vacuum gravity solutions, the Weyl curvature spinor is 
$\Psi_{ABCD} = \frac{1}{S}\Phi_{(AB}\Phi_{CD)}$
for some scalar field $S$ (``zeroth copy'')
and symmetric spinor field $\Phi_{AB}$ (``single copy''), 
where $S$ satisfies a wave equation and $\Phi_{AB}$ 
satisfies Maxwell's equations. 
(We refer to \cite{PR1,PR2} for background on the 2-spinor formalism.)
The relation has been proven for vacuum type D and type N spacetimes 
\cite{typeDDC, typeNDC}.
We shall now show that the integrability conditions of 
complex structures give automatically this sort of relations among 
scalar, Maxwell, and gravitational fields.

Consider a conformal K\"ahler geometry, with conformal factor 
$\Omega$, K\"ahler form 
$\hat{\kappa}_{ab}=\varphi_{AB}\epsilon_{A'B'}$, 
Weyl spinor $\Psi_{ABCD}$ and Ricci spinor $\Phi_{ABA'B'}$.
Then one can show the following identities:
\begin{subequations}\label{KahlerDC}
\begin{align}
 & (\Box + 2\Psi_{2} + R/6) \Omega= 0,  \label{FI} \\
 & {\rm d}\hat\kappa = 0 = {\rm d}^{*}\hat\kappa, 
 \label{KahlerMaxwell}\\
 & \Psi_{ABCD} = 
 \Psi_{2}\Omega^{-4}\varphi_{(AB}\varphi_{CD)}, \label{Weyl} \\
 & \Phi_{ABA'B'} = 
 \Phi_{11} |\Omega|^{-4} \varphi_{AB}\bar\varphi_{A'B'}.
 \label{RicciCK}
\end{align}
\end{subequations}
Eq. \eqref{FI} follows by first noticing that the Lee form $f_{a}$ 
(defined by ${\rm d}\kappa = -2f\wedge\kappa$) is 
$f_{a}=\partial_{a}\log\Omega$, then taking a divergence and 
using the identity $\nabla^{a}f_{a}+f^{a}f_{a}=-(2\Psi_{2}+R/6)$ 
(which can be proved using the Newman-Penrose formalism).
Eq. \eqref{KahlerMaxwell} follows from the 
conformal K\"ahler condition. 
Finally, equations \eqref{Weyl}-\eqref{RicciCK} follow from 
the integrability conditions of a conformal K\"ahler structure, where 
for \eqref{RicciCK} we assume Lorentz signature 
(and that the Ricci tensor satisfies $R(J\cdot,J\cdot)=R(\cdot,\cdot)$).

From \eqref{KahlerDC} we see that any conformal K\"ahler 
geometry combines scalar $\Omega$, 
Maxwell $\hat\kappa$, and gravitational fields in a double copy-like 
structure, without assuming any field equations. 
For Einstein manifolds ($\Phi_{ABA'B'}=0$), Bianchi identities imply
$\Omega=\Psi^{1/3}_{2}$, so we recover the type D double 
copy \cite{typeDDC} (extended to non-trivial cosmological constant).
More generally, all of the conformal K\"ahler examples of the 
previous sections have the structure \eqref{KahlerDC}, so they 
represent double copy relations.
New examples include the Fubini-Study and 
Chen-Teo instantons, but also the whole (non-vacuum) 
Pleba\'nski-Demia\'nski class. 

Furthermore, in the Pleba\'nski-Demia\'nski case, 
the fields $\Omega$ and $\hat\kappa$ 
solve flat spacetime equations.
More precisely, we see from \eqref{ConformalFactor} that 
$\Omega$ is independent of $\{M,N,q_{e},q_{m},\lambda\}$, 
and since the case in which these parameters vanish 
corresponds to Minkowski, we immediately get
$\eta^{ab}\partial_{a}\partial_{b} \Omega = 0$.
In addition, from \eqref{MaxwellField} we see that the K\"ahler form 
$\hat\kappa$ depends only on $a$, so it must solve Maxwell's equations 
in Minkowski. 

The type N double copy \cite{typeNDC} is not included in the 
above construction, but it also arises from the integrability 
of complex structures.
First, consider a Petrov type II spacetime whose repeated 
principal spinor $o_{A}$ satisfies
\begin{align}\label{SFR}
 o^{A}o^{B}\nabla_{AA'}o_{B} = 0.
\end{align}
Eq. \eqref{SFR} is the condition for ``half-integrability'' of a complex 
structure (see \cite[Section 2.4]{Araneda21}).
One can show \cite[Prop. 2.6]{Araneda21}
that there is a scalar $\Omega$ such that 
the Lee form satisfies 
$o^{A}f_{AA'}=o^{A}\nabla_{AA'}\log\Omega$. 
Applying $\iota^{A}\nabla^{A'}_{A}$ to this equation 
(where $o_{A}\iota^{A}=1$), 
after some computations we again find that $\Omega$ satisfies 
\eqref{FI}. 
Notice that $\Omega$ is not unique: we have the freedom to add 
$\Omega \to \Omega+\nu$, where $\nu$ is any function
such that $o^{A}\nabla_{AA'}\nu=0$.

In addition, from \cite[Lemma (7.3.15)]{PR2}, eq. \eqref{SFR} implies 
that there are two complex scalars $z^{i}=(z^{0},z^{1})$ such that 
${\rm d}z^{i} = o_{A}Z^{i}_{A'}{\rm d}x^{AA'}$, for some spinors 
$Z^{i}_{A'}$. It follows that the 2-form ${\rm d}z^{0}\wedge{\rm d}z^{1}$ 
is anti-self-dual and closed, so it is a Maxwell field. 
Note that $F(z^{0},z^{1}){\rm d}z^{0}\wedge{\rm d}z^{1}$ 
is also a Maxwell field for any function $F(z^{0},z^{1})$.

Finally, the conditions on $o_{A}$ imply that there is a scalar $\lambda$
such that $o^{A}\nabla_{AA'}(\lambda o_{B})=0$. 
This leads to $\nabla^{AA'}(\lambda\Omega o_{A})=0$, which in turn implies 
that $\varphi_{A_1...A_n}=\Omega\lambda^{n}o_{A_{1}}....o_{A_{n}}$ 
is a massless free field: $\nabla^{A_1A'_1}\varphi_{A_1...A_n}=0$.
For $n=2$ and $n=4$, we get the spin 1 and 2 fields 
$\varphi_{AB}=\varphi_{2}o_{A}o_{B}$ 
and $\psi_{ABCD}=\psi_{4}o_{A}o_{B}o_{C}o_{D}$, 
with $\varphi_{2}=\Omega\lambda^{2}$ and 
$\psi_{4}=\Omega\lambda^{4}$.
These are related by
\begin{align}\label{typeN}
 \psi_{4} = \frac{1}{\Omega}(\varphi_{2})^{2}.
\end{align}
In the special case in which the spacetime is type N,
$\psi_{ABCD}$ can be chosen to be the Weyl curvature spinor,
and \eqref{typeN} is the type N double copy relation \cite[Eq. (6)]{typeNDC}.
The non-uniqueness noticed in \cite{typeNDC} is due to the freedom 
to include the functions $\nu(z^0,z^1)$, $F(z^{0},z^{1})$ mentioned before.

\section{Discussion}
  
A general expression for the K\"ahler potential $K$ for the class of 
conformal K\"ahler (Lorentzian or Euclidean) 
geometries of the form \eqref{SpecialClass} with two Killing fields 
is given by eq. \eqref{KahlerPotentialGeneral}.
This includes the Pleba\'nski-Demia\'nski and Chen-Teo families. 
The potential $K$ generates not only the metric,
but also the Maxwell field $\hat\kappa$.
Notice that this electromagnetic field is exactly the 
Coulomb field of the Schwarzschild solution, 
or the $\sqrt{\text{Kerr}}$ / magic field of the Kerr solution 
\cite{ArkaniHamed, LyndenBell}.

In $(z,\tilde{z})$ coordinates, 
$K$ is not necessarily expressible in terms of elementary functions.
For example, while a potential for Minkowski is
\begin{align*}
 K(z,\tilde{z}) = 4\log\left[\frac{\tilde{z}^{0} - z^{0}}
 {1+e^{\i (z^{1}-\tilde{z}^{1})}} \right],
\end{align*}
in the Schwarzschild case \eqref{KahlerSchw}, 
$r$ and $(z,\tilde{z})$ are related by  
$r+2M\log(r-2M)=(\tilde{z}^{0} - z^{0})/2$,
which can be solved in terms of the Lambert W function.

Nevertheless,
since the K\"ahler potential contains (locally and up to K\"ahler 
transformations) all the information of the geometry, 
it represents a fully non-linear version of the Debye potentials of 
perturbation theory in GR. 
As such, it is of intrinsic interest for the investigation 
of nonperturbative results for gravitational wave physics, 
further supported by the intriguing manifestation of the 
Newman-Janis shift in $K$ found in this paper, 
and by the fact that, as we showed, 
K\"ahler and complex geometry in GR contain
naturally the known instances of the Weyl double copy.

The general framework and results obtained in this paper 
motivate applications to a variety of exciting problems in different areas of interest. 
In mathematical GR, potential applications include 
the analysis of waves on black hole spacetimes, 
analytic compactifications, 
and possible generalizations of the Chen-Teo instanton.
In gravitational wave science, it would be interesting to make 
explicit connections to modern techniques 
used in scattering amplitudes and quantum field 
theory \cite{Buonanno, Travaglini}. 
The relation between K\"ahler potentials and 
the Newman-Janis shift motivates further investigation 
into the geometric origin of this trick, 
together with connections with its interpretation 
as a generation of intrinsic spin \cite[Chapter X]{Flaherty}, see also 
\cite{ArkaniHamed, Guevara}.
A detailed description of dualities in the Pleb\'anski-Demia\'nski family 
will appear elsewhere \cite{SBToappear}.

\paragraph{Acknowledgements.}
We are grateful to Lars Andersson for his interest in this 
work and for pointing out relations to the Lambert W function 
as well as analytic compactifications.
BA acknowledges the support of the Alexander von Humboldt 
Foundation.

\bibliographystyle{plain}

\begin{thebibliography}{1}

\bibitem{Note1}
The standard definition requires $J$ to be real-valued, but we allow $J$ to be
  complex-valued since we also include metrics with Lorentzian signature. We
  shall continue to use terminology, such as Hermitian, Kaehler, etc., as in
  the Riemannian case.

\end{thebibliography}


\begin{thebibliography}{10}

\bibitem{Penrose76} 
  R.~Penrose,
  {\em Nonlinear Gravitons and Curved Twistor Theory},
  Gen.\ Rel.\ Grav.\  {\bf 7}, 31 (1976).

\bibitem{Newman76}
E.~T.~Newman,
{\em Heaven and Its Properties},
Gen. Rel. Grav. \textbf{7} (1976), 107-111

\bibitem{Plebanski75}
J.~F.~Plebanski,
{\em Some solutions of complex Einstein equations},
J. Math. Phys. \textbf{16} (1975), 2395-2402

\bibitem{PlebanskiRobinson} 
  J.~F.~Plebanski and I.~Robinson,
  {\em Left-Degenerate Vacuum Metrics},
  Phys.\ Rev.\ Lett.\  {\bf 37}, 493 (1976).

\bibitem{NewmanJanis}
E.~T.~Newman and A.~I.~Janis,
{\em Note on the Kerr spinning particle metric},
J. Math. Phys. \textbf{6} (1965), 915-917

\bibitem{Trautman62}
A.~Trautman,
{\em Analytic solutions of Lorentz-invariant linear equations},
Proc. Roy. Soc. Lond. A \textbf{270} (1962), 326-328

\bibitem{Flaherty0}
 E.~J.~Flaherty Jr,  
 {\em An integrable structure for type D spacetimes}, 
 Physics Letters A, 46(6), 391-392 (1974)

\bibitem{Flaherty}
 E.~J.~Flaherty Jr., 
 {\em Hermitian and K\"ahlerian Geometry in Relativity}, 
 Springer Lecture Notes in Physics, Vol. 46 (Springer-Verlag, 
 New York, 1976)

\bibitem{FlahertyHeld} 
 E.~J.~Flaherty Jr., 
 {\em Complex Variables in Relativity}, in
 General Relativity and Gravitation. Vol. 2. One hundred years after the birth of Albert Einstein. 
 Edited by A. Held. New York, NY: Plenum Press, p.207, 1980

\bibitem{Buonanno}
A.~Buonanno, M.~Khalil, D.~O'Connell, R.~Roiban, M.~P.~Solon and M.~Zeng,
{\em Snowmass White Paper: Gravitational Waves and Scattering Amplitudes},
[arXiv:2204.05194 [hep-th]].

\bibitem{ArkaniHamed}
N.~Arkani-Hamed, Y.~t.~Huang and D.~O'Connell,
{\em Kerr black holes as elementary particles},
JHEP \textbf{01} (2020), 046
[arXiv:1906.10100 [hep-th]].

\bibitem{Guevara}
A.~Guevara, B.~Maybee, A.~Ochirov, D.~O'connell and J.~Vines,
{\em A worldsheet for Kerr},
JHEP \textbf{03} (2021), 201
[arXiv:2012.11570 [hep-th]].

\bibitem{PD}
 J.~F.~Pleba\'nski and M.~Demia\'nski,  
 {\em Rotating, charged, and uniformly accelerating mass in general relativity}, Annals of Physics, 98(1), 98-127 (1976)

\bibitem{ChenTeo}
Y.~Chen and E.~Teo,
{\em Five-parameter class of solutions to the vacuum Einstein equations},
Phys. Rev. D \textbf{91} (2015) no.12, 124005
[arXiv:1504.01235 [gr-qc]].

\bibitem{Chow}
D.~D.~K.~Chow and G.~Comp\`ere,
{\em Dyonic AdS black holes in maximal gauged supergravity},
Phys. Rev. D \textbf{89} (2014) no.6, 065003
[arXiv:1311.1204 [hep-th]].

\bibitem{Sen}
A.~Sen,
{\em Rotating charged black hole solution in heterotic string theory},
Phys. Rev. Lett. \textbf{69} (1992), 1006-1009
[arXiv:hep-th/9204046 [hep-th]].

\bibitem{BHDC}
R.~Monteiro, D.~O'Connell and C.~D.~White,
{\em Black holes and the double copy},
JHEP \textbf{12} (2014), 056
[arXiv:1410.0239 [hep-th]].

\bibitem{typeDDC}
A.~Luna, R.~Monteiro, I.~Nicholson and D.~O'Connell,
{\em Type D Spacetimes and the Weyl Double Copy},
Class. Quant. Grav. \textbf{36} (2019), 065003
[arXiv:1810.08183 [hep-th]].

\bibitem{typeNDC}
H.~Godazgar, M.~Godazgar, R.~Monteiro, D.~Peinador Veiga and C.~N.~Pope,
{\em Weyl Double Copy for Gravitational Waves},
Phys. Rev. Lett. \textbf{126} (2021) no.10, 101103
[arXiv:2010.02925 [hep-th]].

\bibitem{JNW}
A.~I.~Janis, E.~T.~Newman and J.~Winicour,
{\em Reality of the Schwarzschild Singularity},
Phys. Rev. Lett. \textbf{20} (1968), 878-880

\bibitem{Podolsky}
J.~B.~Griffiths and J.~Podolsky,
{\em A New look at the Plebanski-Demianski family of solutions},
Int. J. Mod. Phys. D \textbf{15} (2006), 335-370
[arXiv:gr-qc/0511091 [gr-qc]].

\bibitem{SBToappear}
S.~Aksteiner and B.~Araneda, to appear

\bibitem{Newman2002}
E.~T.~Newman,
{\em On a classical, geometric origin of magnetic moments, spin angular momentum and the Dirac gyromagnetic ratio},
Phys. Rev. D \textbf{65} (2002), 104005
[arXiv:gr-qc/0201055 [gr-qc]].

\bibitem{Aksteiner}
S.~Aksteiner and L.~Andersson,
{\em Gravitational Instantons and special geometry},
[arXiv:2112.11863 [gr-qc]].

\bibitem{KLT}
H.~Kawai, D.~C.~Lewellen and S.~H.~H.~Tye,
{\em A Relation Between Tree Amplitudes of Closed and Open Strings},
Nucl. Phys. B \textbf{269} (1986), 1-23

\bibitem{PR1}
R.~Penrose and W.~Rindler,
\newblock {\em Spinors and space-time: Volume 1, Two-spinor calculus and
  relativistic fields}, volume~1.
\newblock Cambridge University Press, 1984.

\bibitem{PR2}
R.~Penrose and W.~Rindler,
\newblock {\em Spinors and space-time: Volume 2, Spinor and twistor methods in
  space-time geometry}, volume~2.
\newblock Cambridge University Press, 1986.

\bibitem{Araneda21}
B.~Araneda, 
{\em Conformal geometry and half-integrable spacetimes},
[arXiv:2110.06167 [gr-qc]].

\bibitem{LyndenBell}
D.~Lynden-Bell,
{\em Electromagnetic magic: The Relativistically rotating disk},
Phys. Rev. D \textbf{70} (2004), 105017
[arXiv:gr-qc/0410109 [gr-qc]].

\bibitem{Travaglini}
G.~Travaglini, A.~Brandhuber, P.~Dorey, T.~McLoughlin, S.~Abreu, Z.~Bern, N.~E.~J.~Bjerrum-Bohr, J.~Bl\"umlein, R.~Britto and J.~J.~M.~Carrasco, \textit{et al.}
{\em The SAGEX Review on Scattering Amplitudes},
[arXiv:2203.13011 [hep-th]].


\end{thebibliography}

\end{document}